\documentclass[twocolumn,superscriptaddress,aps,prl,showpacs]{revtex4-1}
\usepackage{graphicx}
\usepackage{subfigure}
\usepackage{dcolumn}
\usepackage{bm}
\usepackage[usenames]{color}
\usepackage{amsmath}
\usepackage{amssymb}
\usepackage{url} 
\usepackage{slashed}
\usepackage{hyperref}

\def\lsim{\mathrel{\rlap{\lower4pt\hbox{\hskip1pt$\sim$}}
   \raise1pt\hbox{$<$}}}
\def\gsim{\mathrel{\rlap{\lower4pt\hbox{\hskip1pt$\sim$}}
   \raise1pt\hbox{$>$}}}

\begin{document}
\topmargin 0.0001cm
\title{A novel way to search for light dark matter in lepton beam-dump experiments}

\newcommand*{\INFNGE}{Istituto Nazionale di Fisica Nucleare, Sezione di Genova, 16146 Genova, Italy}
\newcommand*{\UNIGE}{Universit\'a degli studi di Genova, 16126 Genova, Italy}
\newcommand*{\INFNCT}{Istituto Nazionale di Fisica Nucleare, Sezione di Catania, 95125 Catania, Italy}
\newcommand*{\UDEA}{Universidad de Antioquia, Instituto de F\'isica, Calle 70 No. 52-21, Medell\'{i}n, Colombia}
\newcommand*{\INFNLNF}{Istituto Nazionale di Fisica Nucleare, Laboratori Nazionali di Frascati, C.P. 13, 00044 Frascati, Italy}
\newcommand*{\INFNRM}{Istituto Nazionale di Fisica Nucleare, Sezione di Roma, 00185 Roma, Italy}
\newcommand*{\UNIRM}{Universit\'a degli studi Roma La Sapienza, 00185 Roma, Italy}
\newcommand*{\Apr}{{A^\prime}}

\author {L.~Marsicano} 
\affiliation{\INFNGE}
\affiliation{\UNIGE}
\author {M.~Battaglieri} 
\affiliation{\INFNGE}
\author {M.~Bond\'i} 
\affiliation{\INFNCT}
\author{C.~D.~R.~Carvajal}
\affiliation{\UDEA}
\author {A.~Celentano} 
\affiliation{\INFNGE}
\author {M.~De~Napoli} 
\affiliation{\INFNCT}
\author {R.~De~Vita} 
\affiliation{\INFNGE}
\author {E.~Nardi} 
\affiliation{\INFNLNF}
\author {M.~Raggi} 
\affiliation{\UNIRM}
\author {P.~Valente} 
\affiliation{\INFNRM}

\date{\today}

\begin{abstract}
A novel mechanism to produce and detect Light Dark Matter in experiments making use of GeV electrons (and positrons) impinging on a thick target (beam-dump) is proposed.
The positron-rich environment produced by the electromagnetic  shower allows to produce an  $\Apr$ via non-resonant ($e^+ + e^- \to \gamma + \Apr$) and resonant  ($e^+ + e^- \to \Apr$) annihilation on   atomic electrons. The latter mechanism, for some selected kinematics, results in a larger  sensitivity with respect to  limits derived by the commonly used $\Apr-strahlung$. This idea, applied to Beam Dump Experiments and {\it active} Beam Dump Experiments pushes down the current limits by an order of magnitude.
\end{abstract}
\maketitle
The Standard Model (SM) of particle physics does not explain some  experimental facts, such as dark matter (DM), neutrino masses, and the cosmological baryon asymmetry. Physics beyond the SM is thus required, which might eventually emerge as a whole new sector containing new particles as well as new interactions. 
These new states do not need to be particularly heavy to have so far escaped detection,  their 
masses could well be within experimental reach, provided they couple sufficiently feebly 
to SM particles. For example, particles with mass below 1 GeV/c$^2$ 
would have easily escaped detection by underground experiments seeking for halo DM, so that complementary searches attempting to cover  this mass region  are  well motivated. 

In a popular scenario, Light Dark Matter (LDM) with mass in the range ($\sim$1 MeV/c$^2$ - 1 GeV/c$^2$)   is charged under a new $U(1)_D$ broken symmetry, whose vector boson mediator $\Apr$ ({\it heavy photon}, also called  {\it dark photon}) is massive. The dark photon can be kinetically mixed with the SM hypercharge field, resulting in SM-DM interaction~\cite{HOLDOM1986196}. The lowest order effective Lagrangian associated to the model, in case of a fermionic $\chi$ DM particle, reads:
\begin{eqnarray}
\mathcal{L}_{eff} &=&
-\frac{1}{4} F'_{\mu\nu}F'^{\mu\nu} 
+ \frac{1}{2} m^2_{\Apr} \Apr_{\mu}{\Apr}^{\mu} 
- \frac{\varepsilon}{2} F'_{\mu\nu}F^{\mu\nu} +
\nonumber
\\
& + & \overline{\chi}\left(i\slashed{D}-m_\chi\right)\chi \;  ,
\end{eqnarray}
where $F'_{\mu\nu}$ is the field strength of the hidden gauge field  $\Apr_{\mu}$, $m_{\Apr}$ and $m_\chi$ the masses of the heavy photon and of the $\chi$ particle respectively, and $F_{\mu\nu}$ the QED photon field strength (at energies much higher than the ones we will consider here,  $F_{\mu\nu}$ should be replaced by the hypercharge field strength). Finally, ${D}_\mu=\partial_\mu - i g_D \Apr_\mu$, with $g_D$ the coupling constant associated to the $U(1)_D$ symmetry. If one makes the natural assumption that $g_D \simeq O(1)$, the kinetic mixing parameter $\varepsilon$ is expected to be in the range of  $\sim 10^{-4} - 10^{-2}$ ($\sim 10^{-6} - 10^{-3}$) if the mixing is generated by one (two)-loops interaction~\cite{Essig:2010ye,DELAGUILA1988633,1126-6708-2008-12-104}.
Depending on the relative mass of the $A'$ and the DM particles, the $\Apr$ can decay only into SM particles ({\it visible } decay) or dominantly to LDM states ({\it invisible } decay). In particular, if $m_\chi < m_\Apr / 2$, and provided that $g_D > \varepsilon e$, the latter scenario dominates. This picture is compatible with the well-motivated hypothesis of DM thermal origin. This assumes that, in the early Universe, DM reaches the thermal equilibrium with the Standard Model particles through an interaction mechanism such as the one described above. The present DM density is therefore a relic remnant of its primordial abundance. This hypothesis provides a relation between the observed DM density and the model parameters, resulting in a clear, predictive target for discovery or falsifiability~\cite{Battaglieri:2017aum}.

LDM received strong attention in recent years, motivating many theoretical and phenomenological studies. It also stimulated the reanalysis and interpretation of old data and promoted  new experimental programs to search both for the $\Apr$ and LDM states~\cite{Alexander:2016aln,Battaglieri:2017aum}. 
In  this context, accelerator-based experiments that make use of a lepton beam of moderate energy ($\sim$ 10 GeV) on a thick target  or a beam-dump show a seizable sensitivity to a wide area of LDM parameter space. Different experimental approaches are possible, each affected by different  backgrounds, and  with specific sensitivity to model parameters.
In   Beam-Dump Experiments (BDE)~\cite{PhysRevD.88.114015}, an intense primary beam is dumped on a passive thick target followed by a significant amount of shielding material. Beside the cascade of SM particles, electrons/positrons stopped in the beam-dump may produce an $\Apr$ decaying to a $\chi/\overline{\chi}$ particles pair, thus resulting in an effective  LDM  secondary beam.
Having a small coupling to ordinary matter, LDM particles propagate through the shielding region to the detector. Scattering  on  electrons and nuclei of the detector active material may result in a detectable signal (in the following, we will only focus on the $\chi-e$ scattering process).
{\it Active} Beam-Dump Experiments (aBDE), instead, use the active dump as a detector, exploiting the missing-energy signature of  produced and  undetected $\chi$ to identify the signal~\cite{PhysRevD.91.094026}. The active dump, detecting the EM shower, allows to measure the energy of individual leptons of a monochromatic beam, provided a beam current low enough to avoid pile-up effects. When an energetic $\Apr$ is produced, its ({\it invisible}) decay products will carry away a significant fraction of the primary beam energy, thus resulting in a visible defect in the energy deposited in the active dump. Signal events are identified when the \textit{missing energy}, defined  as the  difference between the beam energy and the detected energy, exceeds a minimum value $E^{CUT}_{miss}$. A variation of the previous technique is represented by {\it missing momentum experiments}. A thin, passive target with a fast particle tracker are added upstream of the EM calorimeter to measure the momentum of each scattered lepton. Employing a thin target, {\it missing momentum experiments}  are characterized by a lower signal yield,
but the measurement of the momentum, correlated with the energy measured by the calorimeter, allows for a more effective background rejection. {\it Missing momentum experiments}  can also perform a missing-energy search, by ignoring the tracker and using the calorimeter-only information.

\begin{figure}[t!]
\begin{center}
\includegraphics[scale=0.55]{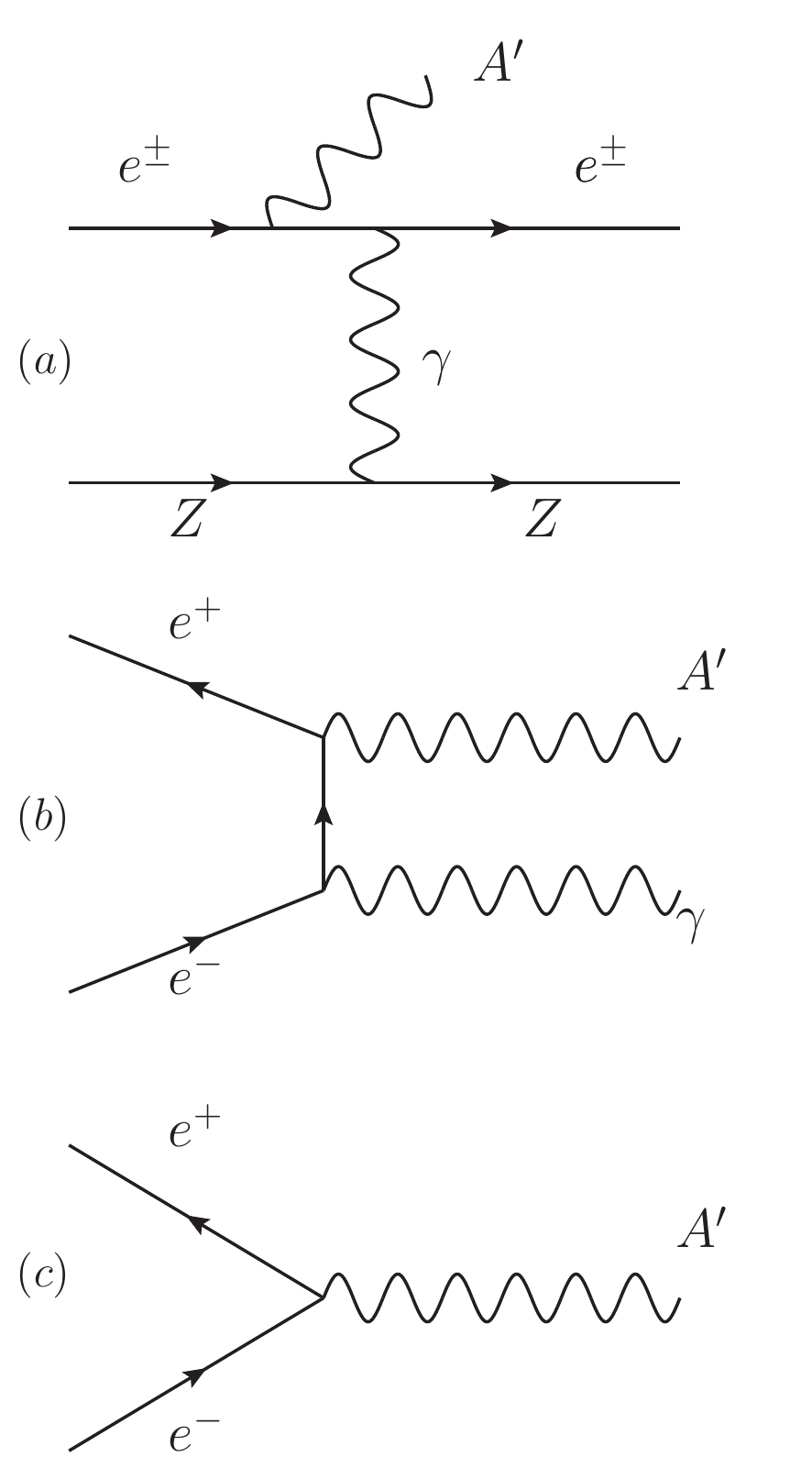}
\caption{Three different $\Apr$ production modes in fixed target lepton beam experiments: 
  $(a)$~$\Apr$-strahlung in $e^-/e^+$-nucleon scattering;
  $(b)$~$\Apr$-strahlung in $e^+e^-$ annihilation; 
  $(c)$~resonant $\Apr$ production in $e^+e^-$ annihilation.}
\label{fig-mech}
 \end{center}
\end{figure}

Dark photons can be generated in collisions of GeV-electrons/positrons with a fixed target by the processes depicted in Fig.~\ref{fig-mech}. For experiments with electron beams, only diagram $(a)$, analogous to ordinary photon bremsstrahlung, has been included in
production estimates for beam-dump setups (we refer to Ref.~\cite{PhysRevD.95.036010} for a critical discussion of limitations of the widely used Weizs\"acker-Williams approximation within this context). The improvement on existing exclusion limits including diagrams $(b)$ and $(c)$  has been discussed in Ref.~\cite{OurPaperPRD} in the context of \textit{visible $\Apr$ } decay.
Regarding fixed target experiments with positron beams, the effect of diagram $(b)$ has been included in the evaluation of the reach for thin-targets setups~\cite{PADME,VEPP3,MMAPS}.
Only recently, the contribution of positron annihilation to the $\Apr$ production
and subsequent {\it visible} decay has been evaluated for a beam-dump experiment~\cite{PADMEeplus}, finding that, for selected kinematics, it provides the dominant contribution. 

In this paper we focus on the effect of positron annihilation in lepton beam-dump experiments searching for LDM through $\Apr$ {\it invisible} decay.
We noticed that in a positron-rich environment produced by the  high-energy electron/positron showering in the dump, contributions from non-resonant ($e^+ + e^- \to \gamma +  \Apr$) and resonant  ($e^+ + e^- \to \Apr$) annihilation can be sizable. These mechanisms significantly enhance the  BDE and  aBDE reach  and  have to be considered for a correct evaluation of the exclusion area in LDM parameters space. 
We calculated the  contribution of positron annihilation for 
past and future electron beam-dump experiments: E137 and LDMX at SLAC \cite{PhysRevD.38.3375,LDMX}, NA64 at CERN \cite{Banerjee:2017hhz}, and BDX at JLab \cite{BDX}. In the context of recent efforts toward a new generation of positron-beam experiments~\cite{jpos}, we also investigated the sensitivity of the same experimental setups replacing the $e^-$ beam with an $e^+$ beam.

\begin{figure}
\vspace{1.cm} 
\includegraphics[width=.5\textwidth]{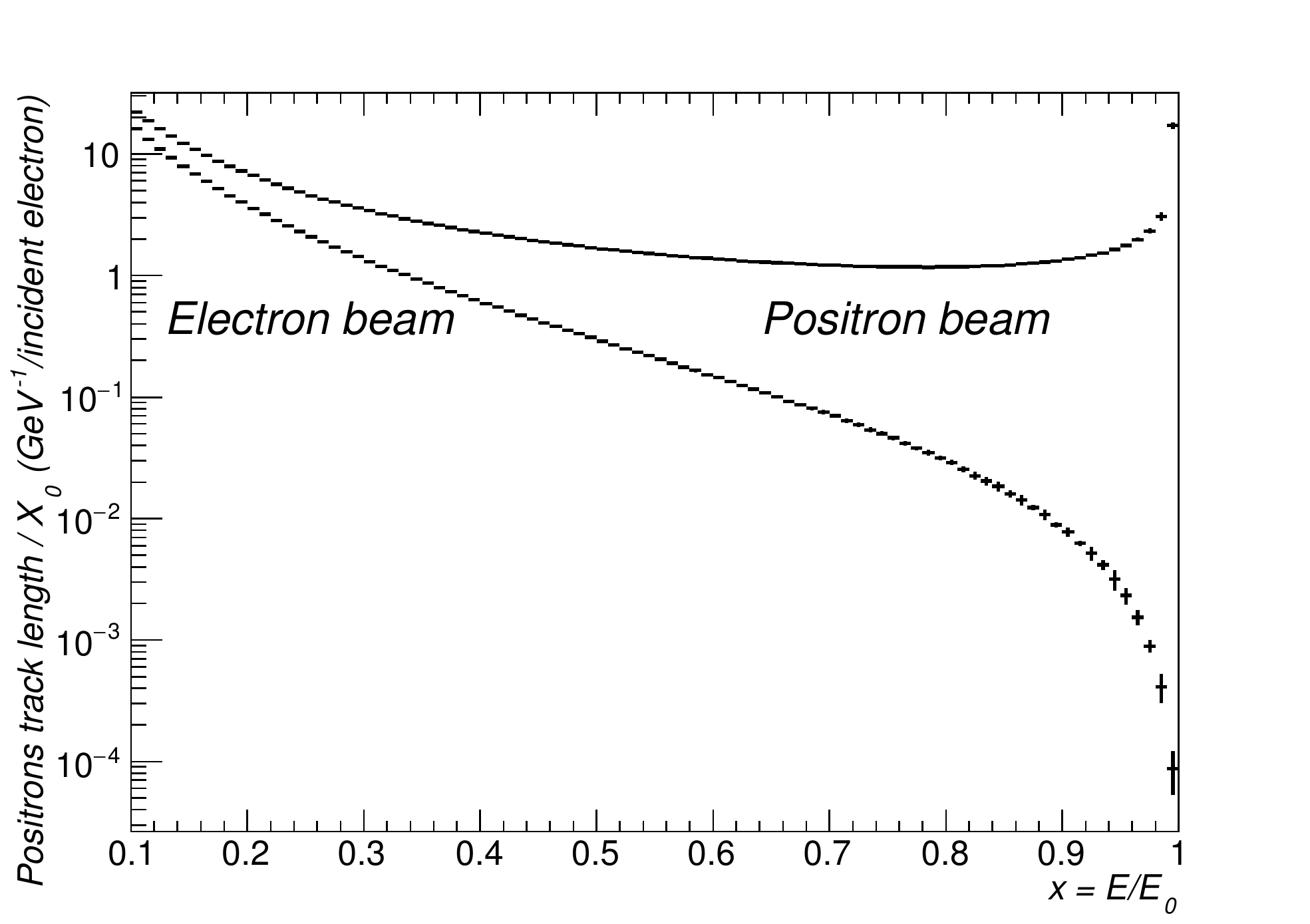}
  \caption{\label{fig-bd-spectra} Differential track-length distribution as a function of $x=E/E_0$ for positrons produced in a thick target by an impinging $e^+/e^-$ beam, normalized to  the radiation length $X_0$. 
}
\end{figure}

We estimated the positron-annihilation contributions using Monte Carlo simulations, as described in details in Ref.~\cite{OurPaperPRD} for $\Apr$ production in the thick target. 
The experimental setups (beam-dump geometry and materials) of aforementioned experiments were implemented in the GEANT4~\cite{AGOSTINELLI2003250} and FLUKA~\cite{BOHLEN2014211}
simulation frameworks. The secondary
positron differential track-length distribution $T_+(E_{e^+},{\Omega}_{e^+})$, as a function of $e^+$ energy $E_{e^+}$ and angles ${\Omega}_{e^+}$, was evaluated in the dump for a primary electron and positron beam. The systematic error associated to the procedure was estimated by comparing the angular-integrated $T_+(E_{e^+})$  obtained within the two simulation frameworks. Agreement within a few percent is observed, validating our approach. Figure~\ref{fig-bd-spectra} shows the ``universal'' $T_+(E_{e^+})$ distribution obtained from the simulations. Data points have been computed from each experimental
setup: $E_{beam}$ = 11 and 20 GeV on an Aluminum target (E137 and BDX), $E_{beam}$ = 4 GeV on a Tungsten target (LDMX), and $E_{beam}$ = 100 GeV on a Lead target (NA64). Points overlap within the error bars. To eliminate the dependence on $E_{beam}$, results are reported as a function of the ``scaling-variable'' $x=E_{e^+}/E_{beam}$. The dependence on the target material is factorized by normalizing each distribution to the corresponding radiation length. This scaling behavior is consistent with the prediction from the analytic function of Tsai~\cite{PhysRev.149.1248}.

%

As a next step, we computed numerically, for kinematically allowed masses, the total number of $\Apr$ ($N_\Apr$) and $\chi/\overline{\chi}$ ($N_{\chi}=N_{\overline{\chi}}$) in the dump. These can be written as:
\begin{equation}\label{eq-prod}
N_\Apr=N_{\chi}=N_{\overline{\chi}} = \frac{N_A}{A} Z \rho \int_{E^R_{min}}^{E_{beam}} dE\int_{4\pi} d\Omega \,\, T_+(E,\Omega)\,\sigma(E) \; ,
\end{equation}
where $A$, $Z$, $\rho$, are, respectively, the beam-dump atomic mass, atomic number, and mass density,  $N_A$ is Avogadro's number, $\sigma(E)$ is the energy-dependent $\Apr$ production cross-section, and $E^R_{min}=\frac{m^2_\Apr}{2m_e}$ is the minimal positron energy required to produce a dark photon with mass $m_{\Apr}$ through positron annihilation on atomic electrons. 
The dark photon decay was assumed to be fully {\it invisible} ($BR(\Apr \rightarrow \chi \overline{\chi})\simeq 1$). The energy and angular distributions of $\chi$ particles were evaluated  numerically by convolving the positrons energy and angular spread in the target with the intrinsic kinematic dependence of  $\Apr$ production and subsequent decay to $\chi$ particles (assumed to be isotropic in the $\Apr$ rest frame).


For BDE, the number of signal events (corresponding to the $\chi-e$ scatterings in the detector) was computed as:
\begin{equation}
N^{s}_{\chi-e}= N_{\chi\overline{\chi}} \, n_e L_{det} \, \sigma^*_{\chi e} \varepsilon_{s} \; ,
\end{equation}
where $N_{\chi\overline{\chi}}$ is the total number of LDM particles ($\chi + \overline{\chi}$) propagating from the beam-dump and impinging on the detector, $L_{det}$ and $n_e$ $(N_{Av}/A \,\rho Z)$ are the detector length and the electron density, respectively, $\varepsilon_{s}$ is the average signal detection efficiency, and $\sigma^*_{\chi e}$ is the total ${\chi-e}$ scattering cross-section integrated over recoil electron energies larger than the detection threshold $E_{thr}$. 
$N_{\chi\overline{\chi}}$ was computed by projecting the  $\chi$  angular distribution in the dump to the detector front-face plane  and measuring the fraction of crossing particles. 
To evaluate $\sigma^*_{\chi e}$ and to determine the energy and angular spectrum of recoiling electrons,  we used the  differential cross-section reported in Ref.~\cite{PhysRevD.88.114015}: 
\begin{equation}
%
\frac{d\sigma_{\chi e}}{dE_R} = 4\pi \alpha \alpha_D \varepsilon^2   m_e \frac{4 m_e m^2_\chi E_R + \left[m^2_\chi +m_e(E-E_R)\right]^2}{(m^2_\Apr+2m_e E_R)^2(m^2_\chi+2m_e E)^2} \;  ,
\end{equation}
where $E$ and $E_R$ are, the $\chi$ and the scattered $e^-$ energies, respectively,  and $\alpha_D =g^2_D/4\pi$.

For aBDE, instead, we computed the number of signal events as the number of $\Apr$ with energy higher than the detector missing-energy cut $E_{miss}^{CUT}$:
\begin{equation}
N^{s}_{\Apr} = \varepsilon_{s} \int^{E_0}_{E^{CUT}_{miss}} N_\Apr (E) dE \; .
\end{equation}

The detection efficiency $\varepsilon_s$ of each experiment we considered  was determined by applying the same selection cuts used in the original analyses.  Further details are given in the following. 

\begingroup
\squeezetable
\begin{table}
\caption{\label{tab:dumps}Main parameters of the E137 and BDX beam-dump experiments.}
\begin{ruledtabular}
\begin{tabular}{llll}
    & \textbf{E137-I} & \textbf{E137-II} & \textbf{BDX}\\
  Beam energy & 20 GeV & 20 GeV & 11 GeV\\
  Electrons on target & $\simeq 6.2\cdot10^{19}$ & $\simeq 1.2\cdot10^{20}$ & $10^{22}$\\
  Target-detector distance & 383 m & 383 m & 25 m \\
  Front-face size & 2x3 m$^2$ & 3x3 m$^2$ & 50x40 cm$^2$\\
  Detector-length $L_{det}$ & 49.5 cm & 13.8 cm & 300 cm\\
 Electrons number density $n_e$ & 7.4$\cdot 10^{23}$ cm$^{-3}$ &  1.5$\cdot 10^{24}$ cm$^{-3}$ &  1.1$\cdot 10^{24}$ cm$^{-3}$\\
Detection threshold  & $\simeq 1$ GeV & $\simeq 1$ GeV & $\simeq 500$ MeV \\
\end{tabular}
\end{ruledtabular}
\end{table}
\endgroup

\begin{figure*}
\includegraphics[width=0.485\textwidth]{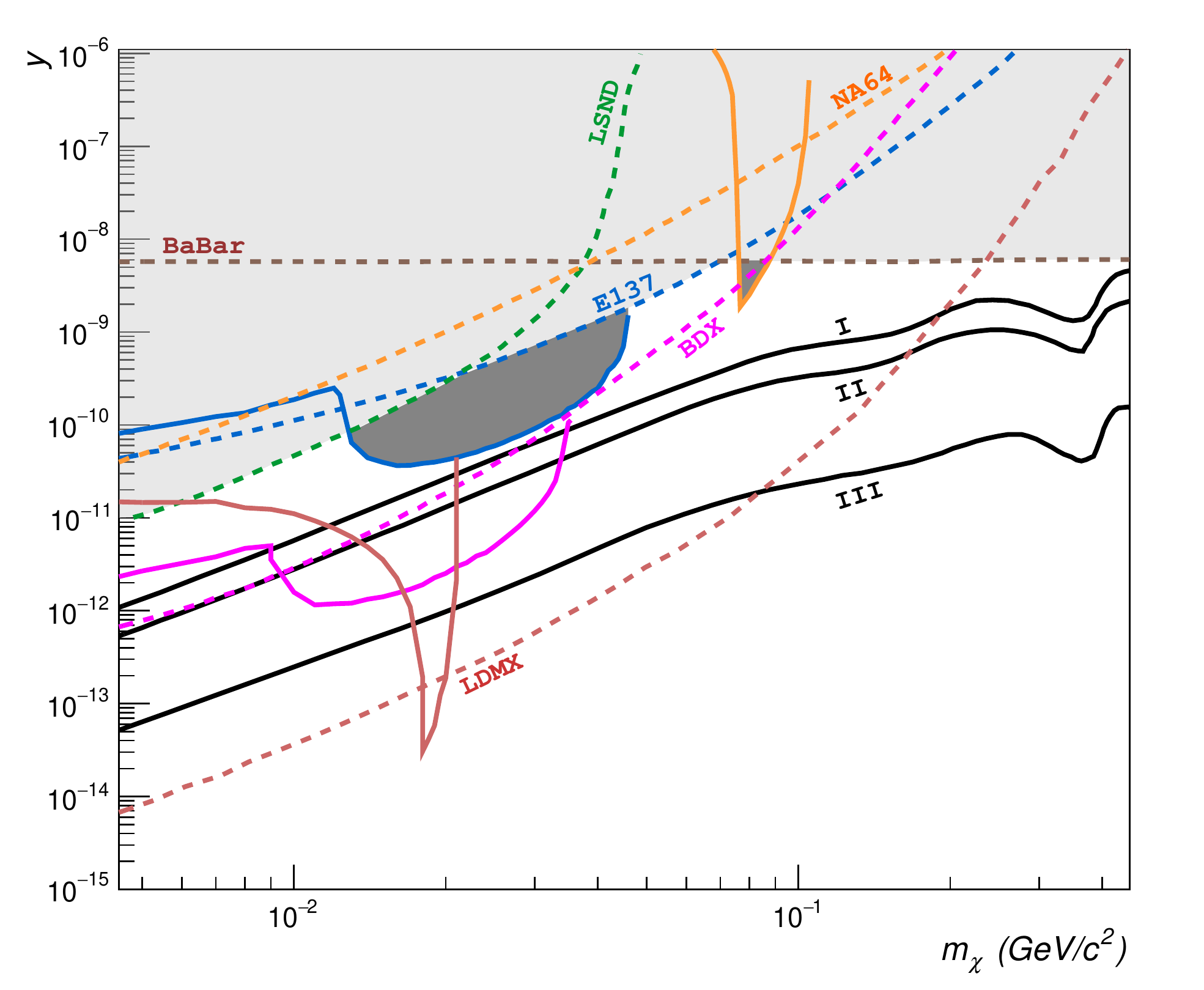}
\quad
\includegraphics[width=0.485\textwidth]{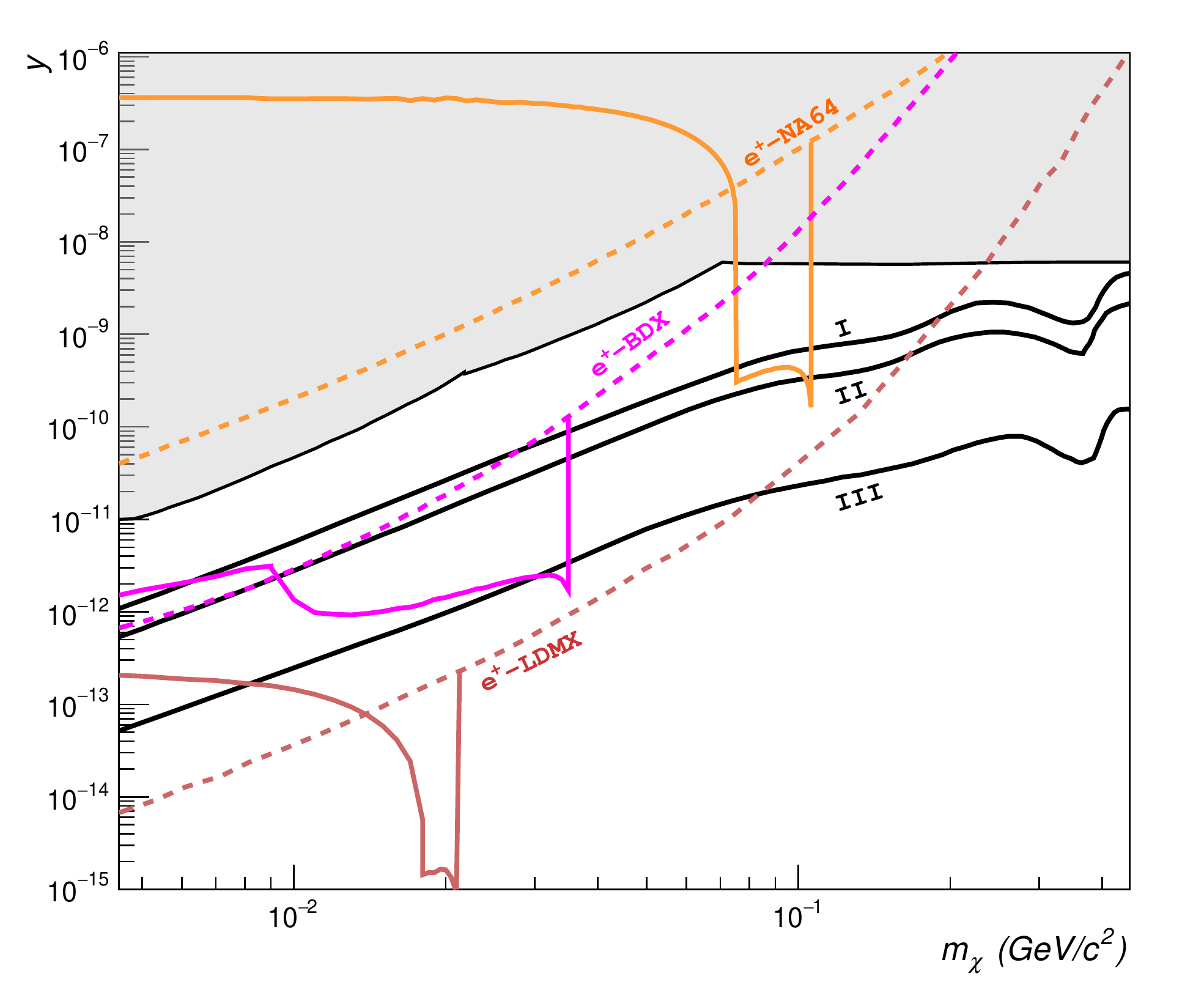}
\caption{
\label{fig:reach}
\textbf{Left:} Continuous lines show exclusion limits at $90\%$ CL for electron BDE and aBDE due to resonant and non-resonant positron annihilation (only). Dashed lines show exclusion limits obtained by considering  $A'-strahlung$ only. The combined exclusion region is shown as filled area: light-gray indicates previous limits (including E137, BaBar~\cite{BABAR} and LSND~\cite{deNiverville:2016rqh});  dark-gray shows the effect of including positron annihilation on existing limits. Different colors correspond to the different experiments: E137 (blue), BDX (magenta), NA64 (yellow), and LDMX (brown).
Limits are given for the parameter $y\equiv \alpha_D \varepsilon^2 \left(\frac{m_\chi}{m_\Apr}\right)^4$ as a function of $m_\chi$. The prescription $\alpha_D=0.5$, $m_\Apr = 3 m_\chi$ is adopted when applicable.
\textbf{Right:} The same as in the Left plot but for possible positron-beam BDE and aBDE. Exclusion limits are derived assuming the same (positron) charge and experimental efficiency quoted for the corresponding $e^-$-beam setup.}
\end{figure*}
\textbf{E137} is a BDE that ran at SLAC in 1980-1982, searching for long-lived neutral objects which might be produced in the electromagnetic shower initiated by 20 GeV electrons in the SLAC Beam Dump East. The main parameters of the experiment are summarized in Tab.~\ref{tab:dumps}.
The detector was an 8-radiation length electromagnetic calorimeter made by a sandwich of a 1 cm plastic scintillator paddles and 1 $X_0$ iron (or aluminum) converters. To satisfy the trigger condition, $\chi$ particles should have scattered in the first 5 layers.
A total charge of $\sim$~30 C was dumped during the live-time of the experiment in two slightly different experimental setups, denoted as ``E137-I'' and ``E137-II'' (see Tab.~\ref{tab:dumps}).
The original data analysis searched for axion-like particles decaying in $e^+$ $e^-$ pairs, requiring a deposited energy in the calorimeter larger than 1 GeV with a track pointing to the beam-dump production vertex. The absence of any signal above threshold, satisfying a tight directionality cut ($\theta_{track} < 30$ mrad), provided stringent limits on axions/photinos. Negative results were used in~\cite{PhysRevLett.113.171802} to provide the most stringent limits in LDM parameters space in the mass range 1 MeV/c$^2$ $<m_\chi < 100$ MeV/c$^2$.

\textbf{BDX} is a BDE planned at JLab that will improve the E137 sensitivity by using the high intensity CEBAF beam~\cite{Freyberger:2015rfv}, running for $\sim$1 year with currents up to $60$ $\mu$A, and positioning the detector closer to the dump. The main parameters of the experiments are reported in Tab.~\ref{tab:dumps}. The detector consists of a $\simeq 0.5$ m$^3$ EM calorimeter made by CsI(Tl) scintillating crystals, surrounded by two active veto layers made by plastic scintillator  for cosmic backgrounds rejection. The average signal efficiency is $\sim15\%$. 
The experiment sensitivity is ultimately limited by the irreducible neutrino background, expected to be  at level of $O($few$)$ events for $10^{22}$ electrons on target (EOT). In case of a negative result, BDX is expected to improve the  E137 exclusion limit by one or two order of magnitudes, depending on the $\chi$ mass.

The \textbf{NA64} experiment is an aBDE making use of   the  100 GeV secondary electron beam at SPS-CERN. The detector consists of an upstream magnetic spectrometer (tracker + bending magnet), followed by an active target, a Shashlik-type EM calorimeter made of lead and plastic-scintillator plates. 
A signal event  is defined as an upstream reconstructed track with less than 50 GeV energy deposited in the EM calorimeter, and no activity in the surrounding veto system or in the hadron-calorimeter installed downstream. 
The corresponding detection efficiency, slightly dependent on $m_\Apr$, is about $50\%$.
NA64 accumulated so far $4.3\cdot10^{10}$ EOT, with no events measured in the signal-search window, and an expected background contribution of 0.12 events. The 90$\%$ CL exclusion limit extracted from the measurement spans from $\varepsilon\simeq 2\cdot10^{-5}$ at $m_\Apr=1$ MeV/c$^2$ to $\varepsilon\simeq 3.6\cdot10^{-2}$ at $m_\Apr=1$ GeV/c$^2$.

\textbf{LDMX} is a missing momentum experiment proposed at SLAC that will use the  LCLS-II 4 GeV $e^-$ beam \cite{Raubenheimer:2018mwt}.
The detector  is made by a silicon tracker  surrounding  a 10$\%$ $X_0$ Tungsten thin-target  to  measure  each  incoming  and  outgoing  electrons individually; a fast and highly hermetic Si-W sampling EM calorimeter, and a hadron-calorimeter used to identify and reject penetrating particles.
In the first phase, LDMX plans to collect $\simeq 10^{14}$ EOT, with an expected sensitivity for a zero background measurement that spans from $\varepsilon\simeq 1.2\cdot10^{-6}$ at $m_\Apr=1$ MeV/c$^2$ to $\varepsilon\simeq 7\cdot10^{-3}$ at $m_\Apr=1$ GeV/c$^2$.
Although LDMX is designed for missing-momentum searches using tracker and the EM calorimeter information, it can also acts as aBDE using the EM calorimeter as an active target with a corresponding energy cut of 1.2 GeV.  

The new exclusion limits at 90$\%$ C.L. obtained considering the positron annihilation mechanisms in the aforementioned experiments are reported in Fig.~\ref{fig:reach}.
In case of E137 and NA64, the limit of the number of signal events derived from published data - zero event measured and an almost null contribution expected from background - is $N_{90\%}=2.3$. For BDX and LDMX, a null background contribution was assumed.
In the left-plot limits derived by including $\Apr-strahlung$ only (dashed lines), positron annihilation only (continuous lines) and the combination of the two on existing limits (filled area) are shown. 
Light-gray area shows the excluded region before this work in the parameters space  
 $y\equiv \alpha_D \varepsilon^2 \left(\frac{m_\chi}{m_\Apr}\right)^4$ vs. $m_\chi$ assuming $\alpha_D=0.5$, $m_\Apr = 3 m_\chi$. The dark-gray area highlights the contribution of the positron annihilation to the previously excluded area. The three continuous black lines represent the thermal relic target for different hypothesis on the LDM nature: elastic and inelastic scalar (I), Majorana fermion (II), and pseudo-Dirac fermion (III). 
For some selected kinematics positron annihilation pushes down by an order of magnitude the exclusion limits. The shape of aBDE lines is related to the high missing-energy threshold. For this class of experiments, the sensitivity at low masses is strongly connected to threshold value, resulting in a sharp dip. For BDE, instead, the threshold effect is less pronounced. Here the energy threshold is usually lower, and the dependence on threshold of the sensitivity for low masses is weaker, resulting in a wider and smoother shape.”
In the right-plot, instead, we report the exclusion limits, in case of null result, by running future BDE and aBDE with positron beams. The total accumulated (positron) charge and the detection efficiency of LDM is assumed to be similar to that of the electron-beam counter-parts. 
Nowadays positron beams with such characteristics are not available. However, proposals to run future experiments at JLab~\cite{jpos} and CERN are currently under discussion. For example, the NA64 experiment could already take data with a positron beam in the LHC run-III~\cite{NA64-comm}.
In the calculation, we assumed the same {\it $\Apr$-strahlung} contribution, at the first-order, for $e^-$ and $e^+$ beams~\cite{Tsai:1973py}. Positron annihilation mechanisms, instead, significantly improve the reach since the secondary positron spectrum is enhanced in case of a positron beam (see Fig.~\ref{fig-bd-spectra}). 

In conclusion, we demonstrated that in a  positron-rich environment, such as the electromagnetic shower produced by the interaction of GeV electrons or positrons with a beam-dump, $e^+$ resonant and non-resonant annihilation annihilation are two LDM production mechanism potentially competitive with the widely considered $\Apr-strahlung$. 
We included the two diagrams in Fig.~\ref{fig-mech}$b$  and~\ref{fig-mech}$c$ in the calculation of the exclusion limits for null results of electron BDE and aBDE obtaining, in some selected kinematics, up to an order of magnitude gain in sensitivity. In particular, the best exclusion limit set by E137, is pushed down by a factor of $\sim$10 for $m_{\chi}$ in the range (20 MeV/c$^2$ - 40 MeV/c$^2$).
These results show that  positron annihilation needs to be included for a correct evaluation of all the LDM exclusion limits obtained from electron beam-dump experiments.
We also speculated about running the same experiments with a genuine positron beam. The significant gain in sensitivity we found suggests to consider positron-beam experiments in future LDM searches.

 C.D.R.C.~acknowledges financial support from 
 COLCIENCIAS (doctoral scholarship 727-2015).
 E.N.~is supported in part by the INFN ``Iniziativa
Specifica'' TAsP-LNF.


\bibliographystyle{apsrev4-1} 
\bibliography{biblio} 

\end{document}